\documentclass[numbered]{trbunofficial}

\usepackage[tiny,rm,pagestyles]{titlesec}
\newpagestyle{trbstyle}{
	\sethead{Loder et al.}{}{\thepage}
}
\usepackage{tikz}
\usepackage{booktabs}
\usepackage{tabularx}
\newcommand{\ml}[1]{``Mobilität.Leben''}
\newcommand{\net}[1]{``9-Euro-Ticket''}
\newcommand{\dt}[1]{``Deutschlandticket''}
\begin{document}
\nolinenumbers
%\pagewiselinenumbers % comment out for final manuscript
\thispagestyle{empty}
% Titlepage
\begin{titlepage}
    \begin{flushleft}
     % Title
        {\bfseries The Mobilität.Leben Study: a Year-Long Mobility-Tracking Panel}\\[0.4cm]
        % Title
  
        \textbf{Allister Loder*}\orcidlink{0000-0003-3102-6564}\\
        \faHome~Chair of Traffic Engineering and Control, Technical University of Munich, Germany\\
        \Letter~Email: \href{mailto:allister.loder@tum.de}{allister.loder@tum.de}\\[0.2cm]

        \textbf{Fabienne Cantner}\orcidlink{0000-0001-7287-5496}\\
        \faHome~Professorship Economics, Technical University of Munich, Germany\\
        \Letter~Email: \href{mailto:fabienne.cantner@tum.de}{fabienne.cantner@tum.de}\\[0.2cm]

        \textbf{Victoria Dahmen}\orcidlink{0009-0004-0392-2526}\\
        \faHome~Chair of Traffic Engineering and Control, Technical University of Munich, Germany\\
        \Letter~Email: \href{mailto:v.dahmen@tum.de}{v.dahmen@tum.de}\\[0.2cm]

        \textbf{Klaus Bogenberger}\orcidlink{0000-0003-3868-9571}\\
        \faHome~Chair of Traffic Engineering and Control, Technical University of Munich, Germany\\
        %DE-80333, Germany\\
        \Letter~Email: \href{mailto:klaus.bogenberger@tum.de}{klaus.bogenberger@tum.de}\\[0.2cm]
        
        * Corresponding author\\[0.3cm]
        % \quickwordcount(main_trb}
	%Word count: 5073 words + \total{table} table(s) $\times$ 250 + \numberofwordsthejournalthinksforthereferences~words for references = 6073~words
 %\totalwordcount~words
		\hfill\break%
	
		\textit{Submitted: }{\AdvanceDate[0]\today}
		\hfill\break%
		
		%Paper submitted for presentation at the 103\textsuperscript{rd} Annual Meeting Transportation Research Board, Washington D.C., January 2024\\
		
    \end{flushleft}
\end{titlepage}

\thispagestyle{empty}
\section*{Abstract}

The \ml~ study investigated travel behavior effects of a natural experiment in Germany. In response to the 2022 cost-of-living crisis, two policy measures to reduce travel costs for the population in June, July, and August 2022 were introduced: a fuel excise tax cut and almost fare-free public transport with the so-called \net~. The announcement of a successor ticket to the \net~, the so-called \dt~, led to the immediate decision to continue the study. The \ml~ study has two periods, the \net~ period and the \dt~ period, and comprises two elements: several questionnaires and a smartphone-based passive waypoint tracking. The entire duration of the study was almost thirteen months. 

In this paper, we report on the study design, the recruitment strategy, the study participation in the survey, and the tracking parts, and we share our experience in conducting such large-scale panel studies. Overall, 3,080 people registered for our study of which 1,420 decided to use the smartphone tracking app. While the relevant questionnaires in both phases have been completed by 818 participants, we have 170 study participants who completed the tracking in both phases and all relevant questionnaires. We find that providing a study compensation increases participation performance. It can be concluded that conducting year-long panel studies is possible, providing rich information on the heterogeneity in travel behavior between and within travelers. 

\vspace{2cm}
\noindent\textit{Keywords}: smartphone survey, GPS tracking, travel behavior, travel diary, public transport pricing

\newpage
%%%%%%%%%

\section{Introduction}

In response to the 2022 cost-of-living crisis in Europe, the German government introduced a three-month fuel excise tax cut and a public transport season ticket for 9 Euro per month, valid on all local and regional services, the so-called \net~. The latter can be considered almost fare-free public transport. The selected months were June, July, and August, i.e., a period characterized by summer holidays. Given the critical role of travel costs in mode choice \cite{hensher_behavioural_1979}, this natural experiment was expected to lead to a modal shift to public transport because the season ticket price cut was so substantial. In addition, many expected that the almost fare-free aspect of the \net~ leads to high levels of induced demand. The success of the \net~ prompted an immediate discussion as well as public and political demand for introducing a successor ticket to the \net~ as soon as possible. This ticket, the so-called \dt~, was finally introduced in May 2023. We set up the \ml~ study to observe this fare policy innovation using questionnaires and GPS tracking to generate travel diaries \cite{nationwide_trb2023,nationwide_report1}.

This behavioral intervention, which could be one of the largest public transport pricing travel behavior experiments, has been studied by many: all of them reported a substantial increase in public transport usage during the validity period of the \net~ and a return to almost pre-ticket levels after the \net~ validity period~\cite{verband_deutscher_verkehrsunternehmen_vdv_abschlussbericht_2022,gaus_9-euro-ticket_2023,nationwide_trb2023,kramer_9-euro-ticket_2022,dietl_9-euro-ticket_2022}. The official and main study was conducted by Association of German Transport Companies, which surveyed more than 200,000 people in Germany \cite{verband_deutscher_verkehrsunternehmen_vdv_abschlussbericht_2022}: around 20\% of all \net~ customers were new customers to public transport. Out of all public transport trips in the months of June, July, and August 2022, 17\% of trips have been shifted from other transport modes, and 10\% of trips have been shifted from the car to public transport, in rural areas, even 13 to 16\%. 16\% of all trips correspond to induced demand. In addition, trip distances increased by 38\% during the \net~ period. Another survey showed that 11\% of all trips conducted during the \net~ period shifted from other modes of transport, while 6\% of all trips were induced \cite{kramer_9-euro-ticket_2022}. Using surveys and GPS tracking, another study concluded that the \net~ did not lead to a change in daily mobility but instead increased leisure travel at the beginning and the end of the ticket's validity period, leaving monetary savings as the main effect of the \net~ \cite{gaus_9-euro-ticket_2023}. Generally, the summer months of June, July, and August usually see less ridership due to the summer holidays, but in 2022 this trend was reversed \cite{dietl_9-euro-ticket_2022}. For the \dt~, first results for the Hamburg metropolitan area suggest that season-ticket ownership increased by 22~\% and ridership increased from 89.3~\% to 95.4~\% of the pre-pandemic levels from April to May 2023 \cite{dey_49-euro-ticket_2023}.

The dominant methodological approach was a cross-sectional survey with online questionnaires and interviews \cite{verband_deutscher_verkehrsunternehmen_vdv_abschlussbericht_2022,kramer_9-euro-ticket_2022,dietl_9-euro-ticket_2022}, and panel studies \cite{francke2022,schlueter2022}. Both are established survey techniques in transport research, also for nationwide household travel surveys \cite{hoogendoorn-lanser_netherlands_2015,eisenmann_deutsches_2018,weis_surveying_2021,swiss_federal_statistical_office_verkehrsverhalten_2017}. The complexity of the fare-policy innovation with the \net~ and the \dt~, however, means that travel behavior changes can result, which are difficult to observe using traditional paper-based or CATI-bases travel diary surveys. For example, one can expect that self-assessments of how individuals travel across transit district zones and district borders can only be answered reliably by individuals with some knowledge of the fare system, not everyone. Here, using GPS tracking can be considered an appropriate supplement or substitute to questionnaires to measure travel behavior and its changes over a long time period. Such tracking-based survey approaches generally work in creating travel diaries \cite{giaimo_will_2010}, and it is beneficial in correcting under- or false reporting in traditional approaches \cite{wolf_elimination_2001,stopher_assessing_2007,bricka_household_2009}. GPS tracking can be done using either GPS loggers, where users can edit or validate entries later on a web browser, or smartphone-based, where users can edit and validate entries directly in the app to create semi-passive travel diaries. Comparative analyses of these approaches suggest that key differences are in the organization of the study (sending out and collecting GPS loggers vs. installing an app) and selection bias, as not everyone has a smartphone or a good command, while both data sources can extract ``meaningful diaries'' \cite{montini_comparison_2015,stopher_smartphone_2018}. In \ml~ we use the smartphone for data collection with an app for the semi-passive travel diary generation together with questionnaires for socio-demographic and related travel behavior attributes. Such survey design has already been successfully tested and implemented in other studies \cite{molloy_mobis_2022,axhausen_empirical_2021,heinonen_e-biking_2023,winkler_timeuse_2022}. Generally, such rich data improves the understanding and modeling of the complex dynamics of individual activity patterns, e.g., as shown in \cite{cirillo_dynamic_2010,islam_unraveling_2012,bhat_incorporating_2016}.

This paper presents our survey method to observe these two natural public transport pricing experiments using questionnaires and smartphone-based semi-passive travel diaries, or GPS tracking. We introduce the study design, the recruiting process, and the study participation in the questionnaires and the tracking. We summarize our experiences in recruiting, user attrition, and data completeness to make recommendations for future studies of similar panel size and duration.

\section{Study design and organization}

\begin{figure}
    \centering
    \resizebox{1\textwidth}{!}{%
    \tikzset{every picture/.style={line width=0.75pt}} %set default line width to 0.75pt        

\begin{tikzpicture}[x=0.75pt,y=0.75pt,yscale=-1,xscale=1]
%uncomment if require: \path (0,523); %set diagram left start at 0, and has height of 523

%Straight Lines [id:da5601394388165042] 
\draw    (447,41) -- (53,41) ;
\draw [shift={(50,41)}, rotate = 360] [fill={rgb, 255:red, 0; green, 0; blue, 0 }  ][line width=0.08]  [draw opacity=0] (8.93,-4.29) -- (0,0) -- (8.93,4.29) -- cycle    ;
\draw [shift={(450,41)}, rotate = 180] [fill={rgb, 255:red, 0; green, 0; blue, 0 }  ][line width=0.08]  [draw opacity=0] (8.93,-4.29) -- (0,0) -- (8.93,4.29) -- cycle    ;
%Shape: Rectangle [id:dp6154037730604429] 
\draw  [draw opacity=0][fill={rgb, 255:red, 120; green, 172; blue, 68 }  ,fill opacity=1 ] (690,230) -- (810,230) -- (810,450) -- (690,450) -- cycle ;
%Straight Lines [id:da5528212873151561] 
\draw    (120,230) -- (120,450) ;
%Straight Lines [id:da011751461743301439] 
\draw    (120,230) -- (207,230) ;
\draw [shift={(210,230)}, rotate = 180] [fill={rgb, 255:red, 0; green, 0; blue, 0 }  ][line width=0.08]  [draw opacity=0] (8.93,-4.29) -- (0,0) -- (8.93,4.29) -- cycle    ;
%Shape: Rectangle [id:dp9098660180008515] 
\draw  [draw opacity=0][fill={rgb, 255:red, 120; green, 172; blue, 68 }  ,fill opacity=1 ] (250,230) -- (370,230) -- (370,450) -- (250,450) -- cycle ;
%Shape: Rectangle [id:dp5329758456063953] 
\draw  [draw opacity=0][fill={rgb, 255:red, 3; green, 144; blue, 214 }  ,fill opacity=1 ] (230,140) -- (750,140) -- (750,220) -- (230,220) -- cycle ;
%Shape: Triangle [id:dp4660238983185083] 
\draw  [draw opacity=0][fill={rgb, 255:red, 254; green, 48; blue, 11 }  ,fill opacity=1 ] (215.06,140.95) -- (200.01,41) -- (229.97,40.97) -- cycle ;
%Shape: Triangle [id:dp9140612008671791] 
\draw  [draw opacity=0][fill={rgb, 255:red, 254; green, 48; blue, 11 }  ,fill opacity=1 ] (395.03,140.01) -- (379.99,40.06) -- (409.92,40.03) -- cycle ;
%Shape: Triangle [id:dp4586368394858422] 
\draw  [draw opacity=0][fill={rgb, 255:red, 254; green, 48; blue, 11 }  ,fill opacity=1 ] (315.01,140.99) -- (299.92,41.03) -- (329.97,41.01) -- cycle ;
%Shape: Triangle [id:dp25805739445685116] 
\draw  [draw opacity=0][fill={rgb, 255:red, 254; green, 48; blue, 11 }  ,fill opacity=1 ] (515.07,140.97) -- (500.04,41.02) -- (529.97,40.99) -- cycle ;
%Shape: Triangle [id:dp8034420633424477] 
\draw  [draw opacity=0][fill={rgb, 255:red, 254; green, 48; blue, 11 }  ,fill opacity=1 ] (665.07,140.97) -- (650.04,41.02) -- (679.97,40.99) -- cycle ;
%Shape: Triangle [id:dp6447255597952439] 
\draw  [draw opacity=0][fill={rgb, 255:red, 254; green, 48; blue, 11 }  ,fill opacity=1 ] (744.99,140.01) -- (729.92,40.06) -- (759.92,40.03) -- cycle ;
%Straight Lines [id:da31642236689032033] 
\draw    (50,450) -- (800,450) ;
%Straight Lines [id:da7688713765815445] 
\draw    (50,300) -- (50,450) ;
%Straight Lines [id:da5203106527608068] 
\draw [color={rgb, 255:red, 0; green, 0; blue, 0 }  ,draw opacity=1 ][line width=2.25]    (70,377.9) -- (110,371.2) ;
%Straight Lines [id:da620528918770137] 
\draw [color={rgb, 255:red, 0; green, 0; blue, 0 }  ,draw opacity=1 ][line width=2.25]    (110,371.2) -- (150,334.5) ;
%Straight Lines [id:da4251967031456354] 
\draw [color={rgb, 255:red, 0; green, 0; blue, 0 }  ,draw opacity=1 ][line width=2.25]    (150,334.5) -- (190,346.8) ;
%Straight Lines [id:da6693422018310069] 
\draw [color={rgb, 255:red, 0; green, 0; blue, 0 }  ,draw opacity=1 ][line width=2.25]    (190,346.8) -- (230,340) ;
%Straight Lines [id:da9289416271727751] 
\draw [color={rgb, 255:red, 0; green, 0; blue, 0 }  ,draw opacity=1 ][line width=2.25]    (230,340) -- (270,350.6) ;
%Straight Lines [id:da09544031274703824] 
\draw [color={rgb, 255:red, 0; green, 0; blue, 0 }  ,draw opacity=1 ][line width=2.25]    (270,350.6) -- (310,363.1) ;
%Straight Lines [id:da06869910853835348] 
\draw [color={rgb, 255:red, 0; green, 0; blue, 0 }  ,draw opacity=1 ][line width=2.25]    (310,363.1) -- (350,373) ;
%Straight Lines [id:da1677504941057164] 
\draw [color={rgb, 255:red, 0; green, 0; blue, 0 }  ,draw opacity=1 ][line width=2.25]    (350,373) -- (390,348.3) ;
%Straight Lines [id:da9856761429449088] 
\draw [color={rgb, 255:red, 0; green, 0; blue, 0 }  ,draw opacity=1 ][line width=2.25]    (390,348.3) -- (430,351.5) ;
%Straight Lines [id:da25068325853564155] 
\draw [color={rgb, 255:red, 0; green, 0; blue, 0 }  ,draw opacity=1 ][line width=2.25]    (430,351.5) -- (470,357.4) ;
%Straight Lines [id:da38973747181450724] 
\draw [color={rgb, 255:red, 0; green, 0; blue, 0 }  ,draw opacity=1 ][line width=2.25]    (470,357.4) -- (510,374.5) ;
%Straight Lines [id:da49273517151295443] 
\draw [color={rgb, 255:red, 0; green, 0; blue, 0 }  ,draw opacity=1 ][line width=2.25]    (510,374.5) -- (550,370.6) ;
%Straight Lines [id:da7184631324308364] 
\draw [color={rgb, 255:red, 0; green, 0; blue, 0 }  ,draw opacity=1 ][line width=2.25]    (550,370.6) -- (590,368) ;
%Straight Lines [id:da47271951324835637] 
\draw [color={rgb, 255:red, 0; green, 0; blue, 0 }  ,draw opacity=1 ][line width=2.25]    (590,368) -- (630,367.5) ;
%Straight Lines [id:da5958425208815721] 
\draw [color={rgb, 255:red, 0; green, 0; blue, 0 }  ,draw opacity=1 ][line width=2.25]    (630,367.5) -- (670,362.3) ;
%Straight Lines [id:da3202875482150651] 
\draw [color={rgb, 255:red, 0; green, 0; blue, 0 }  ,draw opacity=1 ][line width=2.25]    (670,362.3) -- (710,365.5) ;
%Straight Lines [id:da2103987916701342] 
\draw [color={rgb, 255:red, 0; green, 0; blue, 0 }  ,draw opacity=1 ][line width=2.25]    (710,365.5) -- (750,365) ;

%Straight Lines [id:da2800066027197028] 
\draw    (530,450) -- (530,501) ;
%Shape: Rectangle [id:dp09725362987447372] 
\draw  [draw opacity=0][fill={rgb, 255:red, 254; green, 48; blue, 11 }  ,fill opacity=1 ] (0,41) -- (40,41) -- (40,141) -- (0,141) -- cycle ;
%Shape: Rectangle [id:dp29871550580334394] 
\draw  [draw opacity=0][fill={rgb, 255:red, 3; green, 144; blue, 214 }  ,fill opacity=1 ] (0,141) -- (40,141) -- (40,211) -- (0,211) -- cycle ;
%Straight Lines [id:da02505678408314016] 
\draw    (756.92,40.04) -- (453,40.99) ;
\draw [shift={(450,41)}, rotate = 359.82] [fill={rgb, 255:red, 0; green, 0; blue, 0 }  ][line width=0.08]  [draw opacity=0] (8.93,-4.29) -- (0,0) -- (8.93,4.29) -- cycle    ;
\draw [shift={(759.92,40.03)}, rotate = 179.82] [fill={rgb, 255:red, 0; green, 0; blue, 0 }  ][line width=0.08]  [draw opacity=0] (8.93,-4.29) -- (0,0) -- (8.93,4.29) -- cycle    ;
%Straight Lines [id:da30117545532061096] 
\draw    (50,450) -- (50,501) ;
%Straight Lines [id:da28977588902350293] 
\draw  [dash pattern={on 4.5pt off 4.5pt}]  (450,20) -- (450,450) ;

% Text Node
\draw (273,481) node [anchor=north west][inner sep=0.75pt]   [align=left] {2022};
% Text Node
\draw (664,481) node [anchor=north west][inner sep=0.75pt]   [align=left] {2023};
% Text Node
\draw (70.5,452) node [anchor=north] [inner sep=0.75pt]   [align=left] {Jan};
% Text Node
\draw (191,452) node [anchor=north] [inner sep=0.75pt]   [align=left] {Apr};
% Text Node
\draw (310,452) node [anchor=north] [inner sep=0.75pt]   [align=left] {Jul};
% Text Node
\draw (430.67,453) node [anchor=north] [inner sep=0.75pt]   [align=left] {Oct};
% Text Node
\draw (550,452) node [anchor=north] [inner sep=0.75pt]   [align=left] {Jan};
% Text Node
\draw (670.5,452) node [anchor=north] [inner sep=0.75pt]   [align=left] {Apr};
% Text Node
\draw (175,255) node   [align=left] {\begin{minipage}[lt]{74.8pt}\setlength\topsep{0pt}
start of cost-of-living crisis
\end{minipage}};
% Text Node
\draw (311.5,275) node   [align=left] {\begin{minipage}[lt]{74.12pt}\setlength\topsep{0pt}
\begin{center}
9-Euro-Ticket and \\fuel tax cut
\end{center}

\end{minipage}};
% Text Node
\draw (750,275) node   [align=left] {\begin{minipage}[lt]{68pt}\setlength\topsep{0pt}
\begin{center}
Deutschland-ticket
\end{center}

\end{minipage}};
% Text Node
\draw (20,176) node  [rotate=-270] [align=left] {\begin{minipage}[lt]{41.31pt}\setlength\topsep{0pt}
Tracking
\end{minipage}};
% Text Node
\draw (203,43) node [anchor=north west][inner sep=0.75pt]   [align=left] {W1};
% Text Node
\draw (790.5,452) node [anchor=north] [inner sep=0.75pt]   [align=left] {Jul};
% Text Node
\draw (302,43) node [anchor=north west][inner sep=0.75pt]   [align=left] {W2};
% Text Node
\draw (383,43) node [anchor=north west][inner sep=0.75pt]   [align=left] {W3};
% Text Node
\draw (503,43) node [anchor=north west][inner sep=0.75pt]   [align=left] {W4};
% Text Node
\draw (652.04,43) node [anchor=north west][inner sep=0.75pt]   [align=left] {W5};
% Text Node
\draw (733,43) node [anchor=north west][inner sep=0.75pt]   [align=left] {W6};
% Text Node
\draw (48,300) node [anchor=east] [inner sep=0.75pt]   [align=left] {250};
% Text Node
\draw (49,350) node [anchor=east] [inner sep=0.75pt]   [align=left] {200};
% Text Node
\draw (49,400) node [anchor=east] [inner sep=0.75pt]   [align=left] {150};
% Text Node
\draw (48,450) node [anchor=east] [inner sep=0.75pt]   [align=left] {100};
% Text Node
\draw (10.5,375.9) node  [rotate=-270] [align=left] {Price (Euro-Cent/l)};
% Text Node
\draw (221,22) node [anchor=north west][inner sep=0.75pt]   [align=left] {First phase};
% Text Node
\draw (571,22) node [anchor=north west][inner sep=0.75pt]   [align=left] {Second phase};
% Text Node
\draw (12.04,114.38) node [anchor=north west][inner sep=0.75pt]  [rotate=-268.81] [align=left] {Survey};

\end{tikzpicture}
    }%
    \caption{Study design of \ml~.}
    \label{fig:study_design}
\end{figure}
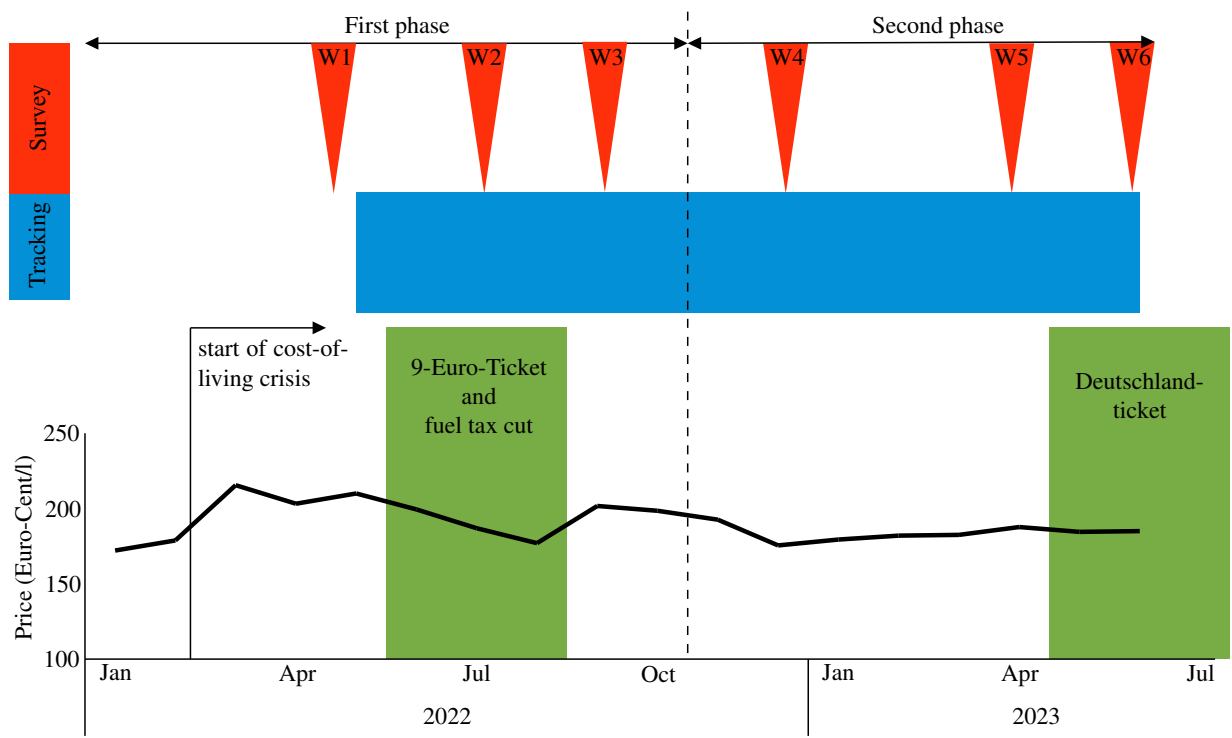

The overall study design of \ml~ is shown in Figure \ref{fig:study_design}. The study has two phases. The first period includes the validity phase of the \net~ and the fuel excise tax cut, and the second phase includes the introduction of the \dt~, the successor ticket to the \net~. We set the end of the first phase and the start of the second phase to 1 November 2022. Figure \ref{fig:study_design} also shows, over time, the development of the average price for petrol in Euro-cent per liter. The substantial price increase that led to the Government's actions can be seen in spring 2022, while the savings from the reduced fuel excise tax are only visible in August, while data suggests that there was no immediate effect at the beginning of June. 

The study comprises two parts: in total, six questionnaires, three in phase one and three in phase two, as well as continuous tracking of participants using a semi-passive travel diary smartphone app. The forth wave was a special questionnaire: given the high cost of energy, it focused on energy consumption and energy conservation measures as well as aimed at motivating participants to keep on tracking although the starting date of the \dt~ and, thus, the total study duration were unknown.

\begin{figure}
    \centering
    \begin{subfigure}[b]{0.3\textwidth}
         \centering
         \includegraphics[width=\textwidth]{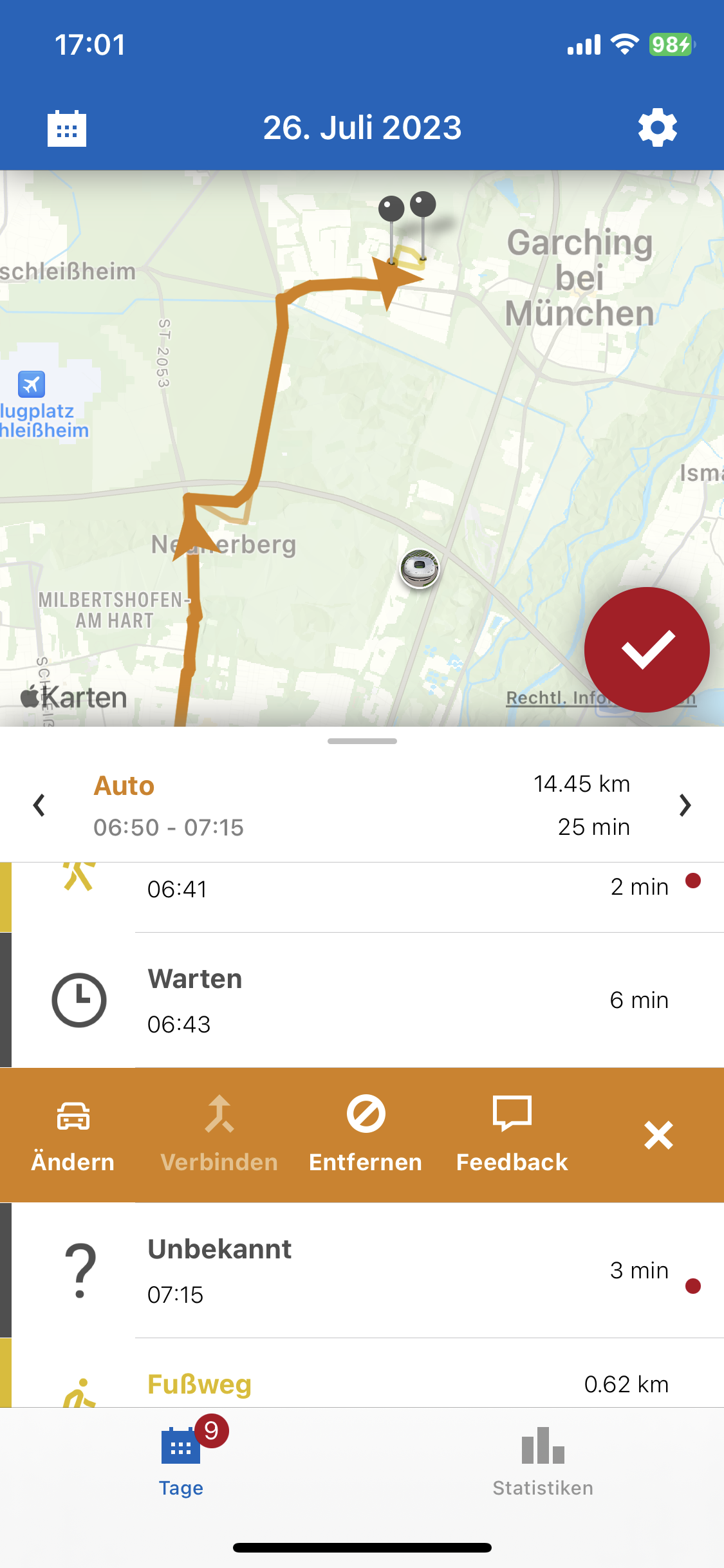}
         \caption{Single travel diary entry}
         \label{fig:single}
     \end{subfigure}
     \begin{subfigure}[b]{0.3\textwidth}
         \centering
         \includegraphics[width=\textwidth]{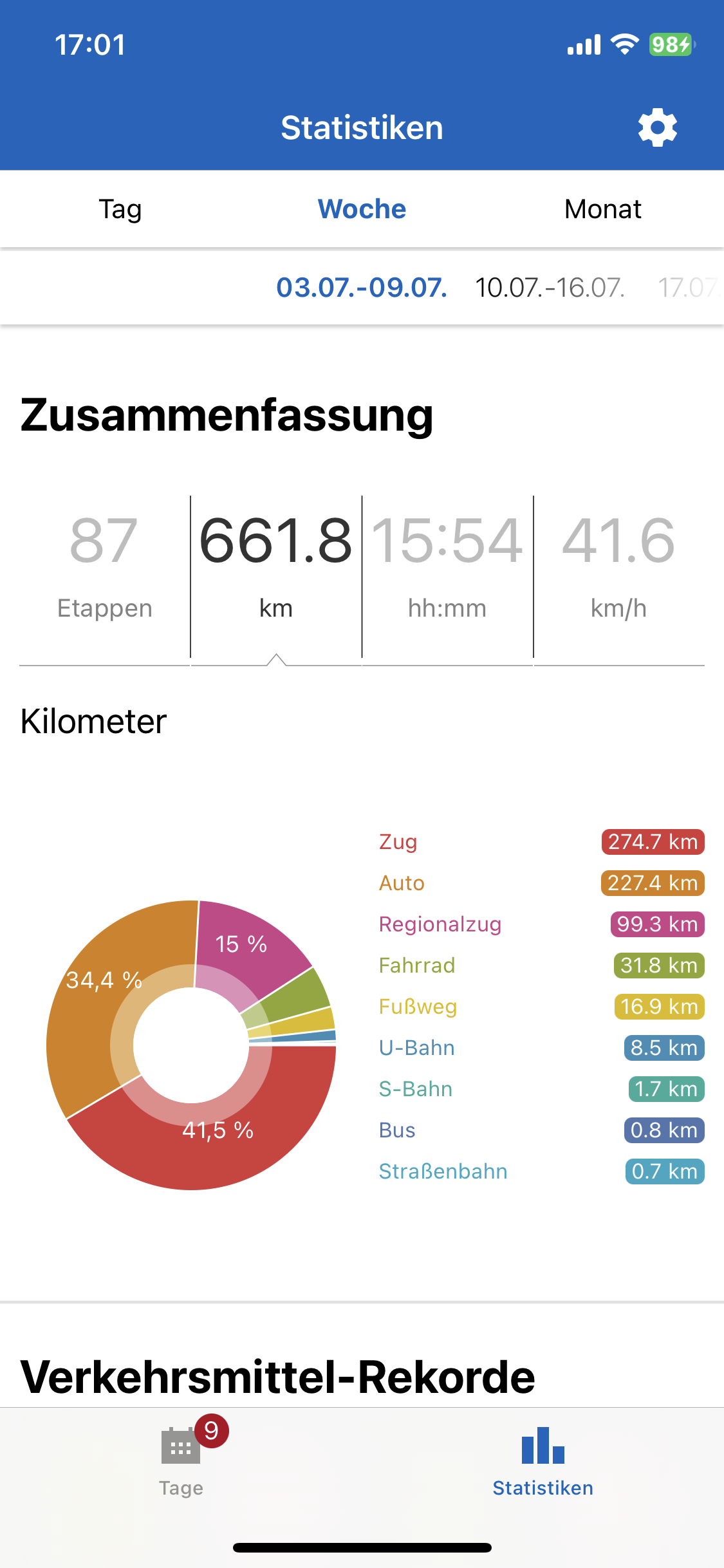}
         \caption{Weekly statistics}
         \label{fig:week}
     \end{subfigure}
    \caption{Screenshots of the smartphone track app with semi-passive travel diary.}
    \label{fig:app}
\end{figure}

Figure \ref{fig:app} shows screenshots of the smartphone app. It displays the travel diary on a map and allows one to edit the entry, comment on entries, and validate entries (Figure \ref{fig:single}. The user sees the following attributes: start- and end times, travel distance, travel distance, and detector or edited mode of transport. The app also features a screen where users can see their personal travel statistics (Figure \ref{fig:week}).

The short planning duration from the announcement of the \net~ on 24 March 2022, passing parliament on 20 May 2022, and its start on 1 June 2022 had implications on the study design of \ml~. The late passing of parliament meant that we had to wait for the public announcement of our study, which in turn reduced the time available to record travel behavior before the introduction of the \net~. Our study was publicly announced on 23 May 2022, and the first tracking measurements were recorded on 25 May 2022.

Further, the general ad-hoc design of the first period of \ml~ as well as the uncertainty of the actual introduction of the \dt~, made a perfect a priori design of the study, its testing, and its communication to study participants impossible. For example, the starting date of the \dt~ was initially announced to be 1 January 2023, but it was postponed several times until the ticket finally started on 1 May 2023. 

The six questionnaires were online questionnaires, each of around ten to fifteen minutes in length. They contained socio-demographic questions, questions on mobility tool ownership and regarding their transport- and energy-related attitudes. Every questionnaire also asked respondents about their travel behavior and their travel behavior changes as a consequence of the \net~ and \dt~. Given the cost-of-living crisis, we also asked respondents in every questionnaire about the impact of this crisis on their households.

Participants received invitations and reminders for the surveys and app via email. This email account also served as the contact point for questions and reporting issues with the smartphone app.

\section{Recruiting}

For the first phase of the \ml~ study, the sample has been recruited in a two-pronged strategy. First, a media campaign in Munich started on 23 May, 2022, aimed at recruiting study participants for the tracking and survey parts. This campaign hat a regional focus on the Munich metropolitan region. Second, a professional agency was tasked to recruit participants across the nation, but only for the survey part of the study. There were two objectives for relying on the professional agency: (i) to obtain nationwide and representative data, (ii) to serve as a backup for our study in case the media-campaign recruitment was unsuccessful given the short time period from the announcement of the \net~ to the start of the study and the \net. People recruited through the professional agency received compensation of about 5~Euros, while participants recruited through the media campaign who completed all surveys of the first phase and provided tracking data until the end of September, i.e., one month after the end of the \net~, received a study compensation of 30~Euros as a voucher.

For the second phase, we asked all participants from the first phase to continue, i.e., the participants recruited through both channels. In addition, a second media campaign was initiated to recruit more participants for the tracking and survey parts. All study participants from the first phase who completed the second phase received a study compensation of 20~Euros as a voucher, while participants recruited through the professional agency were compensated with around 3-4~Euros. 

Overall, this recruitment strategy resulted in the following sample: 3,080 participants registered for the study, 1,706 through the media campaign (1,398 in the first phase and 308 in the second phase), and 1,374 through the professional agency. Not all participants recruited through the media campaign opted to participate in the tracking, leading to 1,420 registered participants (1,112 in the first phase and 308 in the second phase) for the tracking and survey. These participants were invited to install and activate the app. Unfortunately, not all participants were able or willing to activate the tracking app. In the first phase, 186 or 16.7\% of participants never activated the smartphone app, while in the second phase, 77 or 25\% never activated the smartphone app. It is unknown whether the higher activation rate in the first phase is a result of the paid study compensation or the higher interest in research on the \net~. However, evidence from another study suggests that a higher compensation slightly improves study participation \cite{winkler_timeuse_2022}.

Note that an undisclosed number of study participants requested their entire data to be deleted. Consequently, these respondents are not included here. Additionally, as this paper aims to contribute with methodological experiences regarding tracking and survey panel studies, we do not exclude any observations, although some observations are clearly outliers. 

\section{Survey participation}

\begin{figure}
    \centering
    \includegraphics[width=6.3in]{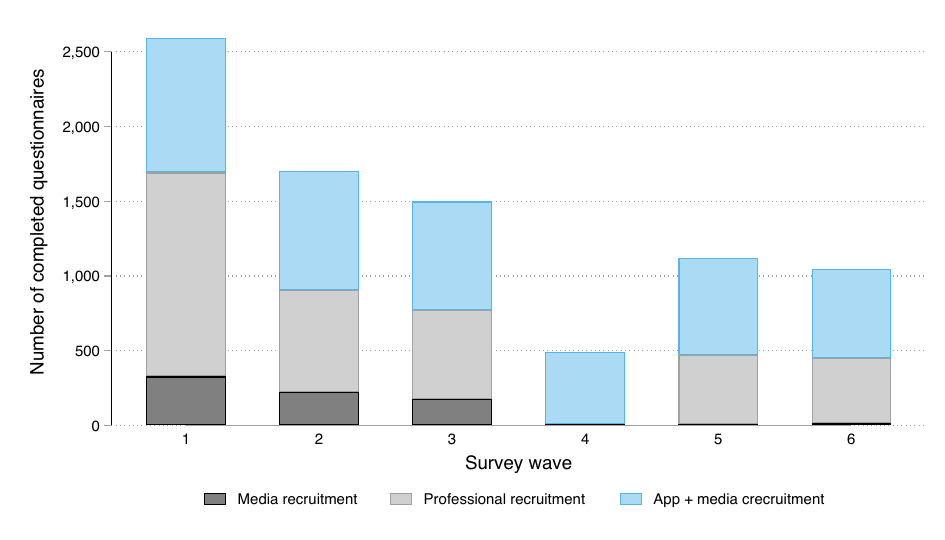}
    \caption{Completion of the six survey waves by recruitment group.}
    \label{fig:waves}
\end{figure}

The participation in each survey is shown in Figure \ref{fig:waves} for three groups: respondents recruited through the professional agency (received study compensation through the agency), respondents recruited through the media campaign and using the smartphone app (only participants recruited in the first phase received a study compensation), and respondents recruited through the media campaign but were not willing or able to activate the smartphone app (received no study compensation). Overall, out of the 3,080 registered participants, 208 participants did not complete any of the six questionnaires; 134 from the first phase (9.6\% of registered participants), 74 from the second phase (24\% of registered participants). The differences can be attributed to a lack of study compensation for participants recruited in the second phase and lower public interest in research on the \dt~ (part of the second phase) than the \net~ (part of the first phase). 

Figure \ref{fig:waves} shows, ignoring the fourth wave, a steady decline in questionnaire completion rates over time even though for the fifth and sixth waves, 308 new study participants were recruited; Figure \ref{fig:waves} also shows a larger decline in completion for the respondents recruited through the professional agency despite receiving a small study compensation. This can be explained by the fact that only around 75\% of first-wave respondents were invited to take part in the second-wave, who provided relevant information for having reliable pre-treatment data, i.e., completed the first wave before the start of the \net~. Note that the fourth wave was not sent to participants recruited through the professional agency as the objective of this wave was to understand energy conservation measures taken by households, to motivate participants to continue tracking and to study participation given that the start date of the \dt~ was unknown.

The assignment of surveys to study phases is shown in Table \ref{tab:survey_participation} together with the completion number per phase. Given the special role of the fourth wave (energy conservation measures, and motivating participants to continue tracking), we consider the completion of both phases with and without the fourth wave. Overall, for the entire sample, we have 27\% of participants who completed all survey waves except the fourth wave, while we have 22\% of all participants recruited through the media campaign who completed all six questionnaires (371 in total). The first phase has been completed by almost 50\% of recruited participants, while only by around 33\% of participants. 

\begin{table}[]
\caption{Completion of survey waves and study phases.}
\label{tab:survey_participation}
\begin{tabularx}{\textwidth}{X|cccccc|ccc}
\toprule
& \multicolumn{6}{c}{Survey   wave} & \multicolumn{2}{c}{Recruitment} & \multicolumn{1}{l}{} \\ \cmidrule{2-7} \cmidrule{8-9}
 &1 & 2 & 3 & 4 & 5 & 6 & Media & Agency & Total \\
 & &  &  &  &  &  & $N=1,706$ & $N=1,374$ & $N=3,080$ \\ 
\toprule
Completed both phases  & x & x & x & x & x & x & 22\% & 0\% & 12\%\\
Completed both phases w/o W4 &x & x & x &  & x & x & 22\% & 32\% & 27\% \\
Completed the first phase &x & x & x &  &  &  & 51\% & 43\% & 47\% \\
Completed the second phase & &  &  &  & x & x & 33\% & 32\% & 33\% \\
\bottomrule
\end{tabularx}
\end{table}

\section{Tracking participation}

\begin{figure}
    \centering
    \includegraphics[width=\textwidth]{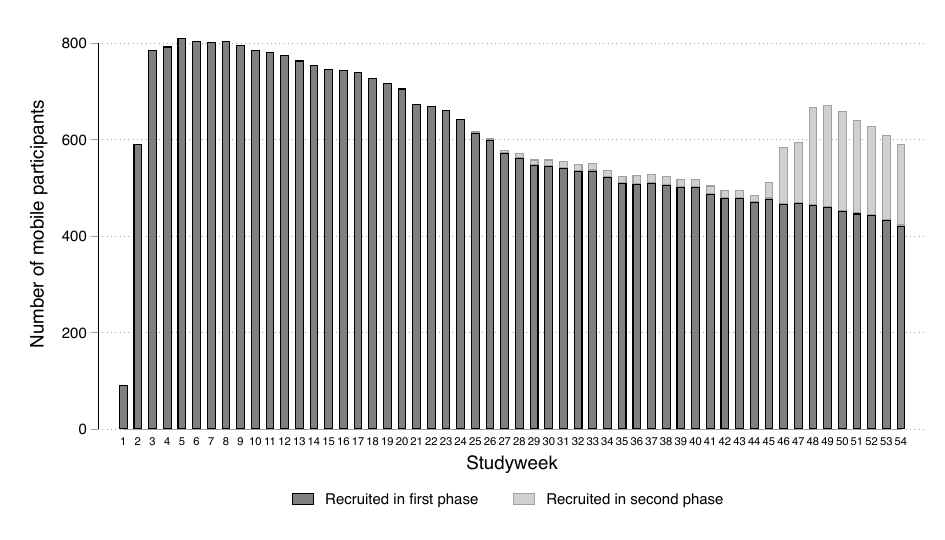}
    \caption{Time series of the number of mobile app users per study week.}
    \label{fig:ts_npart}
\end{figure}

The smartphone app for the semi-passive travel diary generation has been successfully activated and used by 1,137 participants out of 1,420 registered for this study; 909 participants were from the recruitment in the first phase, while 228 were from the recruitment in the second phase. Figure \ref{fig:ts_npart} shows the number of mobile participants per study week, where mobile means that we observed at least one day of travel in the travel diary during that week. The participation peak was in late June to July, i.e., study weeks 4 to 9, with more than 800 participants recording their travel diary in parallel. A second participation peak occurred in the second phase in the end of April and beginning of May, i.e., in study weeks 48 and 49 around the introduction of the \dt~, with almost seven hundred participants providing data in parallel.

Figure \ref{fig:ts_npart} also shows substantial user attrition. For those participants who were, in general, eligible for study compensation, the user attrition rate in the first phase was 0.86\% per week and 1.1\% per week in the second phase. Note that these values are based on the time period defined for receiving the study compensation. For those participants who were not eligible for study compensation as they had been recruited in the second phase, the attrition rate was 4.1\%, around four times the amount of compensated participants; here, the study compensation could be an explanation, but also a different sample with different motives for active participation in the study. 

\begin{figure}
    \centering
    \includegraphics[width=\textwidth]{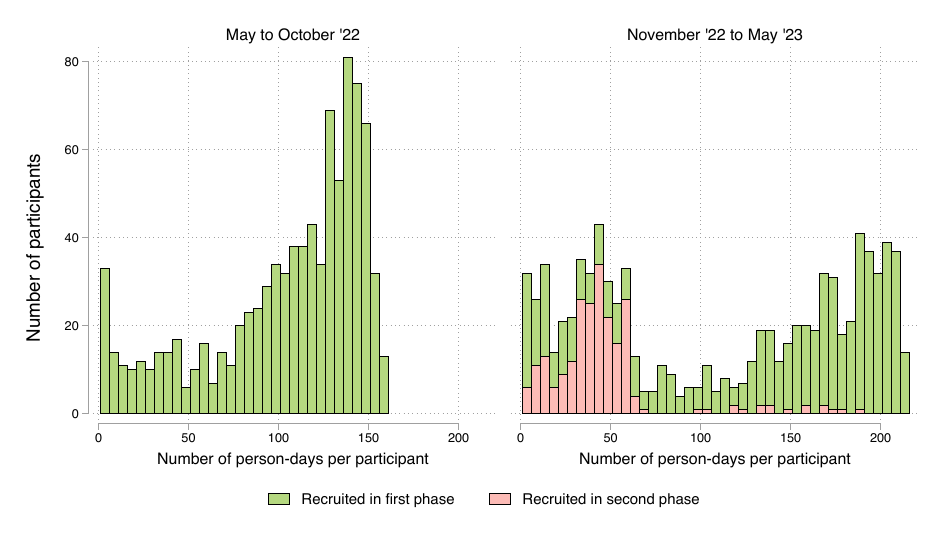}
    \caption{Number of person-days by study phase and recruitment.}
    \label{fig:availble_pdays}
\end{figure}

User participation in such tracking studies are commonly measured in ``person-days'', which counts the number of fully tracked days of a participant. Figure \ref{fig:availble_pdays} shows the available data for both study phases and recruitment time. We find that in both phases, a distribution of person-days skewed towards the total number of days in each phase (note that most participants recruited in the second phase had 60 days as a maximum as they were recruited in March / April 2023). For the entire tracking sample, we find that 13 participants only have a single person-day (1.1\%), 11 have only two person-days (0.9\%), 49 have only seven person-days or less (4.3\%), and 81 have fourteen person-days or less (7.1\%); following \cite{senbil_optimal_2009}, the 14-days threshold can be seen as the minimum amount of person-days required for getting relevant information out of multi-day travel diaries.

\begin{figure}
    \centering
    \includegraphics[width=\textwidth]{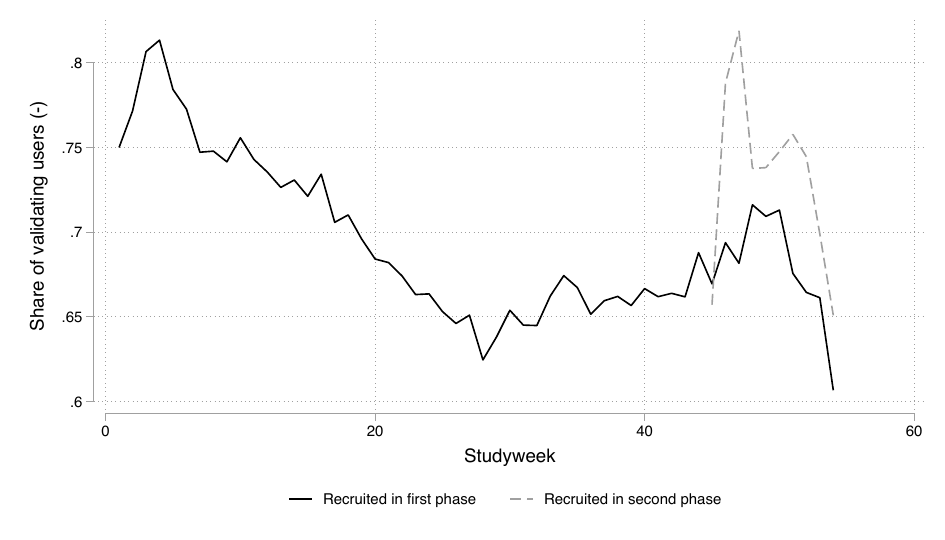}
    \caption{User validation behavior in the smartphone app per study week}
    \label{fig:valibeh}
\end{figure}

The smartphone app offers users the option to edit and validate recorded travel diary entries. An activity is confirmed if it has been labeled as such by pressing the red checkmark button shown in Figure \ref{fig:single}. Users also have the option to confirm entire days at once. This may lead to low-quality validation when inattentive users do not make the effort to check the individual travel diary entries. For this reason, we consider an additional validation metric: a user is considered to be \textit{correcting} if the user actively corrects the draft travel diaries around the time of the recording (we here select a two-week time period). The drafts can be edited in various ways: consecutive tracks of the same mode can be merged, consecutive stays of the same purpose can be merged, and tracks and stays can be removed. In this analysis, we consider users that are correcting, i.e., actively correcting the drafts in the app once in a while, to be \textit{validating} users. Figure \ref{fig:valibeh} shows the resulting validation behavior. 
Generally, we see that around 60\% to 70\% of the sample were actively editing and validating entries in the travel diary, which is consistent with reports of the validation behavior from the MOBIS study \cite{molloy_mobis_2022}. For participants recruited in both phases, validation behavior peaks in the first weeks of app usage before declining steadily. Interestingly, for the participants recruited in the first phase, validation behavior increased towards the start of the \dt~, here, reminders sent via email or the awareness that now relevant data is being collected may have contributed to this increase.

\section{Tracking and survey participation}

Integrating data from the questionnaire and tracking constitutes the core idea of the \ml~ study. Out of the 1,137 participants who reported at least one day of travel through the smartphone app, 26 did not complete any of the three surveys, where, surprisingly, six of them collected entries in their smartphone travel diary for more than 100 days. Considering the full study, we have 170 respondents who completed all survey waves and provided tracking data from before the start of \net~ until the \dt~ was introduced, i.e., around 19\% of the initial active tracking sample that was able to participate in the entire study, i.e., those recruited in the second phase are not considered.

The first study phase was completed by 658 participants, i.e., completing all three survey waves and providing data during and after the \net~ validity period, while 295 also reported travel diary entries before the start of the \net~. Note that due to the short time period between the \net~ bill passing parliament and the official start of the ticket, the recruitment and app onboarding was substantially affected: many study participants registered shortly before or after the start of the ticket and activated their app a couple of days later, which gave them little to no opportunity to collect data before the start. 

The second study phase was completed by 531 participants by completing the fifth and sixth survey waves and collected travel diary data before and after the introduction of the \dt~. This constitutes around 47\% of the initial active tracking sample. 

\section{General experiences and implications}

The \ml~ study is unique in terms of study duration, panel size, and recruitment strategy, especially for the semi-passive travel diary in the smartphone app. The short time period between the \net~ passed parliament and its starting date did not allow for a conventional recruiting strategy, e.g., using direct mail to reach a representative sample of households.  The media campaign used clearly impacted the sample composition and its representativeness (see \cite{nationwide_report1,nationwide_report2} for a discussion and biases). Nevertheless, \ml~ provides valuable experiences and lessons-learned that could be relevant for others planning similar studies.

% General
At the first participation peak in the smartphone app in June 2022 (see Figure \ref{fig:ts_npart}, we received around five to ten emails per week from participants asking for support in the app activation, for setting the correct app parameters, or were reporting errors in the tracking, e.g., gaps in the travel diary. While this amount is not overwhelming, it must be factored into the panel management resources to satisfy participants. Here, future studies could benefit, e.g., from having video documentation on activating and using the travel diary instead of a written FAQ. Given the more or less similar recruiting strategy in both phases, there are substantial differences in registration and app activating numbers as well as study completion and attrition rate, where the sample recruited in the first phase always performed better. This can be attributed to the compensation paid to the sample recruited in the first phase and presumably a higher (altruistic) motivation to contribute to \net~ research, which was widely discussed in public. 

% data quality
Using an integrated survey comprising multiple questionnaires and GPS tracking leads to a complex data structure with quite heterogeneous data quality in the collected travel diaries. Data quality can be influenced by users' validation behavior, but also from the smartphone's operating system \cite{molloy_mobis_2022}, but also the tracking location can matter as speed differences, which factor into mode detection, are relatively close in cities compared to other spatial typologies \cite{marra_developing_2019}. The GPS-based travel diary generation also introduces a source of error. Known issues include a delayed start in the recording and detection of short stays or tracks, which arise from waiting at a level crossing or when walking to a colleague's office. These can, to an extent, be fixed through adequate processing to enhance the quality of the travel diaries. For instance, after such abnormally short tracks are removed, the stays before/after this track can be merged if they are sufficiently close in space and time. Such steps were applied to the dataset used in the following analysis.

Regarding the data quality from the questionnaires, we set 120\,s as a threshold for speeding. Here, we find that 30\% of respondents recruited through the professional agency were speeding, while after excluding them in the following waves, the speeder rate was around 2-4\% throughout the other waves. In the sample recruited through the media campaign, the speeder rate was around 1-2\% throughout all waves. In each survey wave, we asked participants to indicate their average mode usage frequency using the scale from Germany's household travel survey, MiD \cite{mid2018}. While this measure has been established, we experience difficulties in using it for the assessment of an under- or false reporting of trips, but also for measuring a self-reporting bias; hence, recording a full travel diary in the survey would have been more appropriate to assess this difference.  

\section{Discussion and conclusions}

In this paper, we presented our survey method to observe two large-scale nationwide public transport pricing experiments: the \net~ and the \dt~. Our survey method integrates questionnaires with a smartphone app for a semi-passive travel diary collection. We have shown that it is possible to conduct a year-long panel study using such a smartphone-based approach:  following the 80\% success rate in activating the smartphone app after the invitation to download the app, we found an attrition rate of 1\% per week for participants receiving an incentive versus 4\% per week for participants not receiving an incentive. Overall, around 25\% of the invited participants completed the tracking and questionnaires from June 2022 to May 2023.

There are not many similar panel tracking studies in terms of sample size and duration. Most likely, only the MOBIS/COVID study from Switzerland includes more than a thousand participants over almost three years \cite{molloy_mobis_2022} and the AKTA road pricing experiment from Denmark that includes 500 participants over around 100 days \cite{nielsen_akta_2008} are comparable, while the latter uses GPS trackers instead of a semi-passive travel diary. Consequently, as MOBIS/COVID and \ml~ are studies that have been established under exceptional circumstances, the implication for future research is to investigate how such large panel studies can be successfully conducted, made reproducible and valuable, e.g., for household travel surveys, in general. Nevertheless, for such surveys, the study duration must not be in the order of years, but rather in weeks as seen in other projects' study duration, e.g., the three-week period of the ``Lake Geneva Sustainability Monitoring Panel'' \cite{eplf2023} and the six-week period of ``Mobidrive'' \cite{haupt_mobidrive_2001}, or the two-weeks period suggested by \cite{senbil_optimal_2009}, which reduces data collection cost and risk of user attrition substantially. On the methodological side, however, future research must develop methods to obtain the information needed from such week-long semi-passive travel diaries, e.g., methods to separate habitual and regular from irregular travel patterns. As data collection of \ml~ just recently finished, the next steps involve data cleaning, enriching, calculation of observation weights as well as the documentation of the survey (codebook and survey metadata). 

In closing, while the survey method of \ml~ seems powerful to reveal the complexity of spatiotemporal effects associated with the public transport fare policy innovations of the \net~ and the \dt~, the complexity and costs of conducting such a survey are undeniable. Further, many methods to reveal and estimate the effects are still under development, or existing ones require adjustment and testing. Consequently, the still-to-be-answered research question remains, to which degree such a study leads to more information and insights. Considering that artificial intelligence is improving mode and activity detection while the digital literacy of participants is increasing from year to year, both reducing costs and increasing data quality, the probability of a ``yes'' answer to that research question will increase over time.

%%%%%%%%%

\section{Acknowledgment}

Allister Loder acknowledges funding by the Bavarian State Ministry of Science and the Arts in the framework of the bidt Graduate Center for Postdocs. The authors would like to thank the TUM Think Tank at the Munich School of Politics and Public Policy led by Urs Gasser and Markus B. Siewert for their financial and organizational support and the TUM Board of Management for personally supporting the genesis of the project. The authors thank the company MOTIONTAG for handling the app development at utmost priority. The authors would like to thank everyone who supported us in recruiting participants, especially Oliver May-Beckmann and Ulrich Meyer from M Cube and TUM.

%%%%%%%%%%%%%%
\section{Author contributions}

The authors confirm their contribution to the paper as follows: study conception and design: Allister Loder, Klaus Bogenberger; data collection: Allister Loder, Fabienne Cantner, Victoria Dahmen; analysis and interpretation of results: Allister Loder, Fabienne Cantner, Victoria Dahmen, Klaus Bogenberger; draft manuscript preparation: Allister Loder. All authors reviewed the results and approved the final version of the manuscript.

%%%%%%%%%%%%%%%%%%%%%%%%%%%%%

\bibliographystyle{trb}
\setlength{\bibsep}{0pt}
\bibliography{references,extra}

\begin{thebibliography}{34}
\providecommand{\natexlab}[1]{#1}

\bibitem[{Hensher and Stopher(1979)}]{hensher_behavioural_1979}
Hensher, D.~A. and P.~R. Stopher, \emph{Behavioural {Travel} {Modelling}}.
  Routledge, 1979.

\bibitem[{Loder et~al.(2023)Loder, Cantner, Adenaw, and
  Bogenberger}]{nationwide_trb2023}
Loder, A., F.~Cantner, L.~Adenaw, and K.~Bogenberger, The 9 {EUR}-{Ticket} -
  {A} nation-wide experiment: almost fare-free public transport for three
  months in {Germany} - {First} findings. Washington D.C., 2023.

\bibitem[{Loder et~al.(2022)Loder, Cantner, Adenaw, Siewert, Goerg, Lienkamp,
  and Bogenberger}]{nationwide_report1}
Loder, A., F.~Cantner, L.~Adenaw, M.~Siewert, S.~Goerg, M.~Lienkamp, and
  K.~Bogenberger, A nation-wide experiment: fuel tax cuts and almost free
  public transport for three months in {Germany} -- {Report} 1 {Study} design,
  recruiting and participation. \emph{arXiv}, Vol. 2206.00396v1, 2022.

\bibitem[{{Verband Deutscher Verkehrsunternehmen (VDV)} et~al.(2022){Verband
  Deutscher Verkehrsunternehmen (VDV)}, {Deutsche Bahn AG}, and {DB Regio
  AG}}]{verband_deutscher_verkehrsunternehmen_vdv_abschlussbericht_2022}
{Verband Deutscher Verkehrsunternehmen (VDV)}, {Deutsche Bahn AG}, and {DB
  Regio AG}, \emph{Abschlussbericht zur bundesweiten {Marktforschung}}. Verband
  Deutscher Verkehrsunternehmen (VDV), 2022.

\bibitem[{Gaus et~al.(2023)Gaus, Murray, and Link}]{gaus_9-euro-ticket_2023}
Gaus, D., N.~Murray, and H.~Link, 9-{Euro}-{Ticket}: {Niedrigere} {Preise}
  allein stärken {Alltagsmobilität} mit öffentlichen {Verkehrsmitteln}
  nicht. \emph{DIW Wochenbericht}, Vol. 14+15, 2023, pp. 164--171, publisher:
  DIW - Deutsches Institut für Wirtschaftsforschung Version Number: 2.0.

\bibitem[{Krämer et~al.(2022)Krämer, Wilger, and
  Bongaerts}]{kramer_9-euro-ticket_2022}
Krämer, A., G.~Wilger, and R.~Bongaerts, Das 9-{Euro}-{Ticket}: {Erfahrungen},
  {Wirkungsmechanismen} und {Nachfolgeangebot}. \emph{Wirtschaftsdienst}, Vol.
  102, No.~11, 2022, pp. 873--879.

\bibitem[{Dietl and Reinhold(2022)}]{dietl_9-euro-ticket_2022}
Dietl, K. and H.~Reinhold, Das 9-{Euro}-{Ticket}. {Verkehrspolitik} oder
  {Sozialpolitik}? {Eine} {Bewertung} aus {Frankfurter} {Sicht}.
  \emph{Internationales Verkehrswesen}, Vol.~74, No.~4, 2022.

\bibitem[{Dey(2023)}]{dey_49-euro-ticket_2023}
Dey, A., 49-{Euro}-{Ticket} boomt weiter – so viele steigen vom {Auto} um.
  \emph{Hamburger Abendblatt}, 2023.

\bibitem[{Francke(2022)}]{francke2022}
Francke, A., \emph{{Mobilitätserfassung mittels Längsschnitt- und komplexen
  Datenerhebungen: Umfrage zum 9-€-Ticket in drei Wellen}}, 2022.

\bibitem[{Schlueter(2022)}]{schlueter2022}
Schlueter, J., \emph{{9-Euro-Ticket: Auto bleibt dennoch selten stehen}}, 2022,
  \url{https://www.zeit.de/news/2022-08/08/9-euro-ticket-auto-bleibt-dennoch-selten-stehen}.

\bibitem[{Hoogendoorn-Lanser et~al.(2015)Hoogendoorn-Lanser, Schaap, and
  OldeKalter}]{hoogendoorn-lanser_netherlands_2015}
Hoogendoorn-Lanser, S., N.~T. Schaap, and M.-J. OldeKalter, The {Netherlands}
  {Mobility} {Panel}: {An} {Innovative} {Design} {Approach} for {Web}-based
  {Longitudinal} {Travel} {Data} {Collection}. \emph{Transportation Research
  Procedia}, Vol.~11, 2015, pp. 311--329.

\bibitem[{Eisenmann et~al.(2018)Eisenmann, Chlond, Hilgert, von Behren, and
  Vortisch}]{eisenmann_deutsches_2018}
Eisenmann, C., B.~Chlond, T.~Hilgert, S.~von Behren, and P.~Vortisch,
  \emph{Deutsches {Mobilitätspanel} ({MOP}) – {Wissenschaftliche}
  {Begleitung} und {Auswertungen} {Bericht} 2016/2017: {Alltagsmobilität} und
  {Fahrleistung}}. Karlsruher Institut für Technology (KIT) Institut für
  Verkehrswesen, 2018.

\bibitem[{Weis et~al.(2021)Weis, Kowald, Danalet, Schmid, Vrtic, Axhausen, and
  Mathys}]{weis_surveying_2021}
Weis, C., M.~Kowald, A.~Danalet, B.~Schmid, M.~Vrtic, K.~W. Axhausen, and
  N.~Mathys, Surveying and analysing mode and route choices in {Switzerland}
  2010–2015. \emph{Travel Behaviour and Society}, Vol.~22, 2021, pp. 10--21.

\bibitem[{{Swiss Federal Statistical Office} and {Swiss Federal Office for
  Spatial
  Development}(2017)}]{swiss_federal_statistical_office_verkehrsverhalten_2017}
{Swiss Federal Statistical Office} and {Swiss Federal Office for Spatial
  Development}, \emph{Verkehrsverhalten der {Bevölkerung}- {Ergebnisse} des
  {Mikrozensus} {Mobilität} und {Verkehr} 2015}. Bundesamt für Statistik BFS,
  Bundesamt für Raumentwicklung ARE, Neuchâtel, 2017.

\bibitem[{Giaimo et~al.(2010)Giaimo, Anderson, Wargelin, and
  Stopher}]{giaimo_will_2010}
Giaimo, G., R.~Anderson, L.~Wargelin, and P.~Stopher, Will it {Work}?: {Pilot}
  {Results} from {First} {Large}-{Scale} {Global} {Positioning}
  {System}–{Based} {Household} {Travel} {Survey} in the {United} {States}.
  \emph{Transportation Research Record: Journal of the Transportation Research
  Board}, Vol. 2176, No.~1, 2010, pp. 26--34.

\bibitem[{Wolf et~al.(2001)Wolf, Guensler, and Bachman}]{wolf_elimination_2001}
Wolf, J., R.~Guensler, and W.~Bachman, Elimination of the {Travel} {Diary}:
  {Experiment} to {Derive} {Trip} {Purpose} from {Global} {Positioning}
  {System} {Travel} {Data}. \emph{Transportation Research Record: Journal of
  the Transportation Research Board}, Vol. 1768, No.~1, 2001, pp. 125--134.

\bibitem[{Stopher et~al.(2007)Stopher, FitzGerald, and
  Xu}]{stopher_assessing_2007}
Stopher, P., C.~FitzGerald, and M.~Xu, Assessing the accuracy of the {Sydney}
  {Household} {Travel} {Survey} with {GPS}. \emph{Transportation}, Vol.~34,
  No.~6, 2007, pp. 723--741.

\bibitem[{Bricka et~al.(2009)Bricka, Zmud, Wolf, and
  Freedman}]{bricka_household_2009}
Bricka, S., J.~Zmud, J.~Wolf, and J.~Freedman, Household {Travel} {Surveys}
  with {GPS}: {An} {Experiment}. \emph{Transportation Research Record: Journal
  of the Transportation Research Board}, Vol. 2105, No.~1, 2009, pp. 51--56.

\bibitem[{Montini et~al.(2015)Montini, Prost, Schrammel, Rieser-Schüssler, and
  Axhausen}]{montini_comparison_2015}
Montini, L., S.~Prost, J.~Schrammel, N.~Rieser-Schüssler, and K.~W. Axhausen,
  Comparison of {Travel} {Diaries} {Generated} from {Smartphone} {Data} and
  {Dedicated} {GPS} {Devices}. \emph{Transportation Research Procedia},
  Vol.~11, 2015, pp. 227--241.

\bibitem[{Stopher et~al.(2018)Stopher, Daigler, and
  Griffith}]{stopher_smartphone_2018}
Stopher, P.~R., V.~Daigler, and S.~Griffith, Smartphone app versus {GPS}
  {Logger}: {A} comparative study. \emph{Transportation Research Procedia},
  Vol.~32, 2018, pp. 135--145.

\bibitem[{Molloy et~al.(2022)Molloy, Castro, Götschi, Schoeman, Tchervenkov,
  Tomic, Hintermann, and Axhausen}]{molloy_mobis_2022}
Molloy, J., A.~Castro, T.~Götschi, B.~Schoeman, C.~Tchervenkov, U.~Tomic,
  B.~Hintermann, and K.~W. Axhausen, The {MOBIS} dataset: a large {GPS} dataset
  of mobility behaviour in {Switzerland}. \emph{Transportation}, 2022.

\bibitem[{Axhausen et~al.(2021)Axhausen, Molloy, Tchervenkov, Becker,
  Hintermann, Schoeman, Götschi, Castro~Fernández, and
  Tomic}]{axhausen_empirical_2021}
Axhausen, K.~W., J.~Molloy, C.~Tchervenkov, F.~Becker, B.~Hintermann,
  B.~Schoeman, T.~Götschi, A.~Castro~Fernández, and U.~Tomic, \emph{Empirical
  analysis of mobility behavior in the presence of {Pigovian} transport
  pricing}. Bundesamt für Strassen (ASTRA), Eidgenössisches Departement für
  Umwelt, Verkehr, Energie und Kommunikation (UVEK), 2021, volume: 1704.

\bibitem[{Heinonen et~al.(2023)Heinonen, Meyer~de Freitas, Meister, Schwab,
  Roth, Hintermann, Götschi, and Axhausen}]{heinonen_e-biking_2023}
Heinonen, S., L.~Meyer~de Freitas, A.~Meister, L.~Schwab, J.~Roth,
  B.~Hintermann, T.~Götschi, and K.~W. Axhausen, The {E}-biking in
  {Switzerland} ({EBIS}) study: {Methods} and dataset, 2023, paper presented at
  the Swiss Transport Research Conference (STRC).

\bibitem[{Winkler et~al.(2022)Winkler, Meister, Schmid, and
  Axhausen}]{winkler_timeuse_2022}
Winkler, C., A.~Meister, B.~Schmid, and K.~W. Axhausen, {TimeUse}+: {Testing} a
  novel survey for understanding travel, time use, and expenditure behavior.
  \emph{Arbeitsberichte Verkehrs- und Raumplanung}, Vol. 1767, 2022.

\bibitem[{Cirillo and Axhausen(2010)}]{cirillo_dynamic_2010}
Cirillo, C. and K.~W. Axhausen, Dynamic model of activity-type choice and
  scheduling. \emph{Transportation}, Vol.~37, No.~1, 2010, pp. 15--38.

\bibitem[{Islam and Habib(2012)}]{islam_unraveling_2012}
Islam, M.~T. and K.~M.~N. Habib, Unraveling the relationship between trip
  chaining and mode choice: evidence from a multi-week travel diary.
  \emph{Transportation Planning and Technology}, Vol.~35, No.~4, 2012, pp.
  409--426.

\bibitem[{Bhat et~al.(2016)Bhat, Astroza, Bhat, and
  Nagel}]{bhat_incorporating_2016}
Bhat, C.~R., S.~Astroza, A.~C. Bhat, and K.~Nagel, Incorporating a multiple
  discrete-continuous outcome in the generalized heterogeneous data model:
  {Application} to residential self-selection effects analysis in an activity
  time-use behavior model. \emph{Transportation Research Part B:
  Methodological}, Vol.~91, 2016, pp. 52--76.

\bibitem[{Senbil and Kitamura(2009)}]{senbil_optimal_2009}
Senbil, M. and R.~Kitamura, The optimal duration for a travel survey -
  empirical observations. \emph{IATSS Research}, Vol.~33, No.~2, 2009, pp.
  54--61.

\bibitem[{Cantner et~al.(2022)Cantner, Nachtigall, Hamm, Cadavid, Adenaw,
  Loder, Siewert, Goerg, Lienkamp, and Bogenberger}]{nationwide_report2}
Cantner, F., N.~Nachtigall, L.~S. Hamm, A.~Cadavid, L.~Adenaw, A.~Loder, M.~B.
  Siewert, S.~Goerg, M.~Lienkamp, and K.~Bogenberger, A nation-wide experiment:
  fuel tax cuts and almost free public transport for three months in {Germany}
  -- {Report} 2 {First} wave results. \emph{arXiv}, 2022.

\bibitem[{Marra et~al.(2019)Marra, Becker, Axhausen, and
  Corman}]{marra_developing_2019}
Marra, A.~D., H.~Becker, K.~W. Axhausen, and F.~Corman, Developing a passive
  {GPS} tracking system to study long-term travel behavior.
  \emph{Transportation Research Part C: Emerging Technologies}, Vol. 104, 2019,
  pp. 348--368.

\bibitem[{mid(2018)}]{mid2018}
\emph{Mobilit{\"a}t in Deutschland Ergebnisbericht}. Bonn, 2018.

\bibitem[{Nielsen and Sørensen(2008)}]{nielsen_akta_2008}
Nielsen, O.~A. and M.~V. Sørensen, The {AKTA} {Road} {Pricing} {Experiment} in
  {Copenhagen}. In \emph{Road {Pricing}, the {Economy} and the {Environment}}
  (C.~Jensen-Butler, B.~Sloth, M.~M. Larsen, B.~Madsen, and O.~A. Nielsen,
  eds.), Springer Berlin Heidelberg, Berlin, Heidelberg, 2008, pp. 93--109.

\bibitem[{{EPFL}(2023)}]{eplf2023}
{EPFL}, \emph{{Survey: Lake Geneva Sustainability Monitoring Panel}}, 2023,
  {\url{https://www.epfl.ch/labs/lasur/ongoing-surveys/survey-lake-geneva-sustainability-monitoring-panel-2/}}.

\bibitem[{Haupt et~al.(2001)Haupt, Zimmermann, Kübel, Beckmann, Rindfüser,
  Wehmeier, Beckmann, Düsterwald, Axhausen, Schönfelder, König, Schlich,
  Simma, and Fraschini}]{haupt_mobidrive_2001}
Haupt, T., A.~Zimmermann, A.~Kübel, K.~J. Beckmann, G.~Rindfüser,
  T.~Wehmeier, J.~Beckmann, M.~Düsterwald, K.~W. Axhausen, S.~Schönfelder,
  A.~König, R.~Schlich, A.~Simma, and E.~Fraschini, \emph{Mobidrive -
  {Dynamik} und {Routinen} im {Verkehrsverhalten}}, 2001, abschlussbericht an
  das Bundesministerium für Forschung und Technologie.

\end{thebibliography}

% End line numbering
\nolinenumbers

\end{document}